\documentclass[aps,prl,twocolumn,amsmath,amsfonts,floatfix,letterpaper]{revtex4}

\usepackage{graphicx}
\usepackage{ifpdf}

\newcommand{\ee}{\mathrm{e}}
\newcommand{\ii}{\mathrm{i}}
\newcommand{\dd}{\mathrm{d}}

\newcommand{\beq}[1]{\begin{equation}\label{#1}}
\newcommand{\eeq}{\end{equation}}
\newcommand{\refeq}[1]{Eq.~(\ref{#1})}

\newcommand{\beqm}[1]{\begin{multline}\label{#1}}

\newcommand{\punc}[1]{\,{\text{#1}}}
\newcommand{\sub}[1]{_{\text{#1}}}

\newcommand{\Ham}{\mathcal{H}}

\newcommand{\nd}{^{\phantom{\dagger}}}
\newcommand{\ns}{^{\phantom{*}}}

\newcommand{\xh}{\hat{x}}
\newcommand{\yh}{\hat{y}}

\newcommand{\xhv}{\hat{\boldsymbol{x}}}
\newcommand{\yhv}{\hat{\boldsymbol{y}}}

\newcommand{\Tx}{\mathcal{T}_x}
\newcommand{\Ty}{\mathcal{T}_y}
\newcommand{\Txy}{\mathcal{T}_{x,y}}

\newcommand{\kv}{\boldsymbol{k}}

\newcommand{\zerov}{\boldsymbol{0}}

\newcommand{\X}{\boldsymbol{X}}
\newcommand{\Y}{\boldsymbol{Y}}

\newcommand{\Mm}{\mathbf{M}}

\newcommand{\alphav}{\boldsymbol{\alpha}}

\newcommand{\BZ}{\mathfrak{B}}

\newcommand{\at}{\tilde{a}}

\newcommand{\dkv}{\frac{\dd^2 \kv}{(2\pi)^2}}

\newcommand{\putinscaledfigure}[1]{\begin{center}\includegraphics[width=\columnwidth]{#1}\end{center}}

\usepackage{color}

\begin{document}

\title{Interacting Hofstadter spectrum of atoms in an artificial gauge field}

\author{Stephen Powell}
\author{Ryan Barnett}
\author{Rajdeep Sensarma}
\author{Sankar Das Sarma}
\affiliation{Joint Quantum Institute and Condensed Matter Theory Center, Department of Physics, University of Maryland, College Park, MD 20742, USA}

\begin{abstract}
Motivated by experimental advances in the synthesis of gauge potentials for ultracold atoms, we consider the superfluid phase of interacting bosons on a square lattice in the presence of a magnetic field. We show that superfluid order implies spatial symmetry breaking, and predict clear signatures of many-body effects in time-of-flight measurements. By developing a Bogoliubov expansion based on the exact Hofstadter spectrum, we find the dispersion of the quasiparticle modes within the superfluid phase, and describe the consequences for Bragg spectroscopy measurements. The theory also provides an estimate of the critical interaction strength at the transition to the Mott insulator phase.
\end{abstract}


\maketitle


The spectrum of a particle on a tight-binding lattice in the presence of a magnetic field \cite{Harper,Zak,Hofstadter} is a problem that is simple to state but has surprisingly rich phenomena. In the infinite system, it is sensitively dependent on the precise value of $\alpha$, the magnetic flux per plaquette of the lattice (measured in units of the flux quantum). For rational $\alpha = p/q$ ($p$ and $q$ coprime), the spectrum splits into $q$ bands, and each state is $q$-fold degenerate. When the density of states as a function of energy is plotted against $\alpha$, the resulting `Hofstadter butterfly' \cite{Hofstadter} has a fractal structure.

Recent work using cold atomic gases has raised the possibility of observing the Hofstadter spectrum directly in experiment \cite{Jaksch,Mueller,Sorensen,Gerbier}. Two distinct approaches to producing effective magnetic fields for neutral atoms have been proposed and implemented. In a rotating optical lattice \cite{Schweikhard,Williams,Cooper}, the Coriolis force simulates the magnetic Lorentz force, while external lasers applied to a stationary lattice can imprint motion-dependent phases to synthesize a gauge potential \cite{Spielman,Lin}.

Previous theoretical work on this system \cite{Cooper} has considered vortex pinning for both shallow \cite{Reijnders} and deep \cite{Goldbaum} lattice potentials, and the transition from the superfluid to a Mott insulator \cite{Niemeyer,Oktel,Goldbaum,Umucalilar,Sinha}. In this work, we introduce an alternative theoretical approach to the system, using Bogoliubov theory \cite{Bogoliubov,AGD} to provide a controlled expansion for the superfluid phase, in terms of both thermal and quantum fluctuations. We consider the case with both $\alpha$ and mean density $\rho$ (particles per site) of order unity, which is of direct relevance to experiment. To the best of our knowledge, this is the first theoretical study of the superfluid phase in this parameter regime using the Bose-Hubbard model. Starting from this microscopic description, and including the exact Hofstadter spectrum, we calculate the condensate configuration based on a minimization of the on-site interactions, without resorting to a phenomenological Ginzburg-Landau functional. This aspect is similar to earlier work on frustrated Josephson junction arrays \cite{Teitel,Polini,Kasamatsu}, and related models of bosons and fermions in optical lattices \cite{Lim,Zhai}.

Besides describing the real-space condensate configuration, which demonstrates the intricate interplay of the magnetic vortices and the external lattice potential, our approach predicts features that should be directly observable in time-of-flight images. In particular, we show that the superfluid necessarily breaks translation symmetry, resulting in extra peaks in time-of-flight images, clearly distinguishing this case from that with zero flux \cite{Greiner} and also with noninteracting bosons \cite{Gerbier}.

We then go beyond mean-field theory to calculate the spectrum of Bogoliubov quasiparticles, which inherits the complex structure of the Hofstadter spectrum, while also exhibiting the characteristic features of the Bose-Einstein condensate, including a linearly dispersing phonon mode. This spectrum is accessible in experiments using Bragg or lattice-modulation spectroscopy, and we make predictions for the appropriate response functions.

The approach that we describe is valid deep within the superfluid phase, in contrast to previous theory \cite{Niemeyer,Oktel,Goldbaum,Umucalilar,Sinha}, and is directly relevant to experiments. Within our theory, we calculate the depletion of the condensate, due to both thermal and quantum fluctuations, providing limits on the validity of the approach and also approximate boundaries for the superfluid phase. The range of applicability can be extended by taking into account higher-order terms in the Bogoliubov expansion, which give interactions between the quasiparticles.


We begin with the Hamiltonian for a single-band Bose-Hubbard model on a two-dimensional square lattice, in the presence of a uniform magnetic flux of $\alpha$ per plaquette \cite{Jaksch,Jaksch1},
\begin{multline}
\label{Ham}
\!\!\!\!\!\Ham = -t \!\sum_j\! \left( b^\dagger_{j+\xh} b\nd_j + b^\dagger_{j-\xh} b\nd_j + \omega^{x_j} b^\dagger_{j+\yh} b\nd_j + \omega^{-x_j} b^\dagger_{j-\yh} b\nd_j \right)\\
{} + \sum_j \left[ \frac{U}{2} n_j(n_j - 1) - \mu n_j \right]
\punc{,}
\end{multline}
where $b_j$ and $n_j$ are the annihilation and number operators on site $j$ with coordinates $(x_j,y_j)$; $t$, $U$, and $\mu$ are the hopping strength, Hubbard interaction, and chemical potential; and $\omega = \ee^{2\pi \ii \alpha}$. We use the Landau gauge, with the vector potential vanishing on links in the $x$ direction, which is both theoretically convenient and best suited to experiments using an optically-induced gauge potential \cite{Spielman,Lin}. For rotating lattices \cite{Williams}, the symmetric gauge is more natural, requiring a straightforward gauge transformation to be applied.

For simplicity, we will treat an infinite spatially-uniform system and assume rational $\alpha$ throughout. The main effect of the finite trap size is to wash out the small-scale fractal structure of the Hofstadter butterfly \cite{Hofstadter,Jaksch}, so our results should be valid for either rational or irrational $\alpha$. We assume that other effects of the trapping potential can be incorporated using a local-density approximation.

Assuming rational $\alpha = p/q$, one has $\omega^q = 1$, so the magnetic unit cell spans $q$ sites in the $x$ direction, and the Brillouin zone (BZ) is correspondingly reduced. It is helpful to introduce the magnetic translation group \cite{Zak}, with primitive translation operators $\Txy$ obeying $\Tx b_j = b_{j+\xh} \Tx \omega^{-y_j}$ and $\Ty b_j = b_{j+\yh} \Ty$. The phase factor in the former expression, resulting from using the dynamical rather than canonical momentum, gives $[\Txy,\Ham] = 0$. In momentum space, $\Tx$ shifts by $\Y = 2\pi\alpha \yhv$, while the hopping term in $\Ham$ mixes states separated by $\X = 2\pi\alpha \xhv$.

This leads to a noninteracting spectrum \cite{Hofstadter} consisting of $q$ bands (labeled by $\gamma$), within which each state is $q$-fold degenerate. The single-particle states $\psi_{\gamma n}(\kv)$ are given by linear combinations of momenta $\kv + n \X$ for $n = 0,\ldots,q-1$. We define annihilation operators $a_{\kv \ell \gamma}$, for crystal momentum $\kv + \ell \Y$, where $\ell = 0,\ldots,q-1$ and $\kv$ is the momentum referred to the doubly-reduced BZ $\BZ$, $-\frac{\pi}{q}\le k_x,k_y < \frac{\pi}{q}$. The energy associated with these single particle states is $\epsilon_{\gamma}(\kv)$, independent of $\ell$. (The case $\alpha = \frac{1}{3}$ is shown by the dashed line in Figure~\ref{FigDispersions13} below.)

Written using the operators $a_{\kv\ell\gamma}$, the kinetic term in $\Ham$ is diagonal, while the interaction term can be expressed using coefficients $u$ depending on four sets of $\kv$, $\ell$, and $\gamma$. These coefficients are constrained by symmetries \cite{Balents} such as translation, which enforces conservation of crystal momentum.


Within the superfluid phase, the $\mathrm{U}(1)$ phase symmetry is spontaneously broken and the boson annihilation operator has a nonzero expectation value $\langle b_i \rangle$. Following Bogoliubov \cite{Bogoliubov,AGD}, we use the ansatz
\beq{Ansatz}
a_{\kv \ell \gamma} = A_{\ell \gamma}\, (2\pi)^2 \delta^2(\kv) + \at_{\kv \ell \gamma}\punc{,}
\eeq
where the c-numbers $A_{\ell\gamma}$ give the condensate order parameter and the operators $\at_{\kv \ell \gamma}$ describe fluctuations out of the condensate. The condensate can contain contributions from all bands $\gamma$ and from crystal momenta $\ell \Y$ with nonzero $\ell$, and so also breaks symmetry under $\Ty$ (but not $\Ty^q$). Note that, at least within our mean-field theory, translational symmetry is simultaneously broken at the condensation transition \cite{Polini,Balents}.

Our approach is to expand the Hamiltonian in powers of the operators $\at_{\kv \ell \gamma}$, which in physical terms is an expansion in fluctuations, both thermal and quantum. The lowest order in this expansion contains no operators and leads to mean-field theory, while the terms with a single operator cancel. The quadratic terms describe Bogoliubov excitations above the condensate, and higher orders give interactions between these modes.

The mean-field energy density is
\begin{multline}
\label{MFT}
h_0 = \sum_{\ell,\gamma} A^*_{\ell \gamma} A\ns_{\ell\gamma} [\epsilon_\gamma(\zerov) - \mu] \\
{}+ \sum_{\{\ell\},\{\gamma\}}\bar{u}_{\{\ell\},\{\gamma\}} A^*_{\ell_1\gamma_1}A^*_{\ell_2\gamma_2} A\ns_{\ell_3\gamma_3}A\ns_{\ell_4\gamma_4}\punc{,}
\end{multline}
where $\bar{u}$ is the interaction coefficient $u$ with all momenta $\kv_{1\ldots4} = \zerov$ (and a factor of volume removed). The quantity $h_0$ should be minimized with respect to $A_{\ell \gamma}$ to give the mean-field condensate wavefunction in momentum space. The corresponding real-space quantity, $\langle b_j \rangle = \sum_{\ell n} \omega^{n x_j + \ell y_j - n\ell} \sum_\gamma \psi_{\gamma n}(\zerov) A_{\ell\gamma}$, is a function of $x_j$ and $y_j$ modulo $q$; the ansatz of \refeq{Ansatz} implies a real-space unit cell of $q \times q$ sites. (The density pattern minimizing the mean-field energy for $\alpha = \frac{1}{3}$, which consists of diagonal stripes, is shown in the right inset of Figure~\ref{FigDispersions13} below.) The minimization of $h_0$ is therefore equivalent to minimization with respect to the real-space condensate configuration. The latter perspective \cite{Reijnders,Goldbaum} is more appropriate in the continuum, whereas here the lattice provides a strong pinning potential that simplifies the momentum-space approach.

The $q\times q$ real-space unit cell implies a reduced BZ, with observable consequences in time-of-flight images, where Bragg peaks are expected at points corresponding to momenta $\kv$ equal to reciprocal lattice vectors \cite{Greiner}. (Note that the natural gauge for time-of-flight measurements depends on the experimental procedure. We use the Landau gauge appropriate to current experiments with artificial gauge potentials \cite{Spielman,Lin}.) The peak for $\kv = n \X + \ell \Y$ has intensity $|\sum_\gamma A_{\ell \gamma} \psi_{\gamma n}|^2$, multiplied by an envelope due to the momentum-space Wannier wavefunction. We find that on-site interactions generally favor condensate configurations with intensity (but not phase) independent of $\ell$, implying that the only $\ell$ dependence of the peak intensity results from the Wannier envelope. Significantly, the peaks with $\ell \neq 0$ result from spontaneous translational symmetry breaking and are not expected in the noninteracting case \cite{Gerbier}, so time-of-flight measurements give a clear signature of many-body effects.

Like the single-particle states, the condensate configurations are at least $q$-fold degenerate, corresponding to different symmetry-related spatial orderings. This allows for the possibility of real-space domain formation, which would not affect time-of-flight images and would likely require more sophisticated {\it in situ} probes to confirm.


Minimization of the mean-field energy $h_0$ causes the terms in $\Ham$ linear in the operators $\at_{\kv \ell \gamma}$ to vanish. Combining the creation and annihilation operators into $\alphav^\dagger_{\kv\ell\gamma} = (\at^\dagger_{\kv\ell\gamma},\at\nd_{-\kv\ell\gamma})$ and `flattening' this into a $2q^2$-component vector for each $\kv$, the quadratic terms can be written as
\beq{QuadraticHam}
\Ham^{(2)} = \frac{1}{2}\int_{\kv\in\BZ}\!\dkv \, \alphav^\dagger_{\kv} \Mm(\kv)  \alphav\nd_{\kv} + \Ham\sub{c}\punc{,}
\eeq
where $\Ham\sub{c}$ is a constant. Similarly to $h_0$, the matrix $\Mm(\kv)$ has contributions from both the kinetic and interaction energy; the latter are self-energy terms for the quasiparticles due to scattering with bosons in the condensate. They include `anomalous' processes in which a pair of condensed particles scatter from each other into an excited state and the reverse process where they return to the condensate, resulting in quadratic terms that do not conserve the number of $\at_{\kv \ell \gamma}$ quanta \cite{Bogoliubov,AGD}.

In the limit $\kv \rightarrow \zerov$, $\Mm(\zerov)$ is the Hessian matrix for the mean-field energy $h_0$ (with respect to variations of $A_{\ell\gamma}$ and its conjugate) and so is a nonnegative-definite matrix, with a single vanishing eigenvalue corresponding to the $\mathrm{U}(1)$ symmetry of $h_0$. For nonzero $\kv$, all eigenvalues of $\Mm(\kv)$ are strictly positive.

To find the quasiparticle modes, one must diagonalize the quadratic form in \refeq{QuadraticHam} using a Bogoliubov transformation, which for bosons amounts to finding the symplectic transformation that diagonalizes $\Mm(\kv)$ \cite{Blaizot}. The quadratic part of the Hamiltonian then becomes
\beq{QuadHam2}
\Ham^{(2)} = \int_{\kv \in \BZ}\!\dkv \sum_{\zeta} \xi_{\kv\zeta} d^\dagger_{\kv\zeta} d\nd_{\kv\zeta} + \Ham\sub{c}'\punc{,}
\eeq
where $d_{\kv\zeta}$ is the annihilation operator for the Bogoliubov mode labeled by $\zeta = 1,\ldots,q^2$, with energy $\xi_{\kv\zeta}$. While the Bogoliubov quasiparticles have definite momentum referred to $\BZ$, they cannot be labeled by $\ell$, because the condensate configuration breaks the symmetry under translation $\Ty$.

The constant $\Ham\sub{c}'$ in \refeq{QuadHam2} gives the zero-point correction to $h_0$ which, along with the contribution from thermally excited quasiparticles, can in principle change the global minimum of the free energy and hence the most stable condensate configuration.

Using the fact that $\Mm(\kv)$ is positive definite for $\kv \neq \zerov$, one can show that the mode energies $\xi_{\kv\zeta}$ are all real and positive \cite{Blaizot}. At $\kv = \zerov$, one of the eigenvalues vanishes, giving the Goldstone mode resulting from the broken phase rotation symmetry. This mode has linear dispersion near $\kv = \zerov$, with phonon velocity independent of the direction of $\kv$, while the remaining $q^2 - 1$ modes have a nonzero gap and quadratic dispersion. Figure~\ref{FigDispersions13} shows the dispersions $\xi_{\kv\zeta}$ for $\alpha = \frac{1}{3}$, found by minimizing $h_0$ and diagonalizing $\Mm(\kv)$ numerically. For all $\alpha$, there are $q^2$ modes, including one Goldstone mode; here only the lowest $6$ are shown, for clarity.
\begin{figure}
\putinscaledfigure{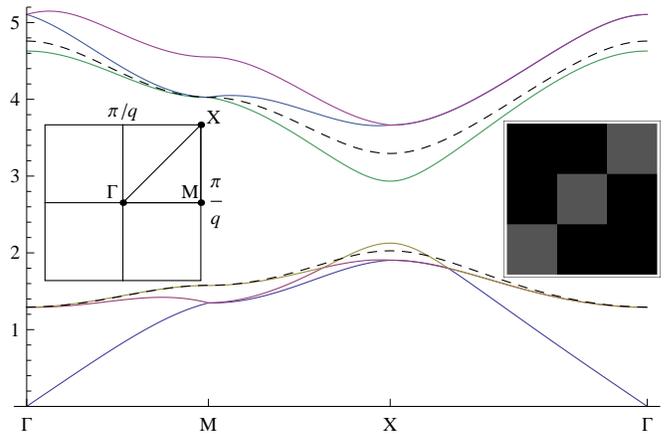}
\caption{\label{FigDispersions13}(Color online) Quasiparticle dispersion (solid lines) and noninteracting single-particle dispersion (dashed), both in units of hopping $t$, for $\alpha = \frac{p}{q} = \frac{1}{3}$. The dispersions are plotted along a path in the reduced Brillouin zone $\BZ$ shown in the left inset. In the interacting case, $U = 2t$, the mean density is $\rho = 1$, and the real-space density pattern is as shown in the right inset (one unit cell of $q \times q$ sites; black squares are sites with higher density than gray squares). In both cases, there are $q^2 = 9$ modes, of which only the lowest $6$ are shown. For $U = 0$, the modes are $q$-fold degenerate and have been shifted vertically by an arbitrary choice of chemical potential.}
\end{figure}


The Bogoliubov ansatz, \refeq{Ansatz}, and expansion in powers of operators is in principle exact, with the higher-order terms leading to interactions between the quasiparticles. The present approximation, truncating the series at quadratic order, is valid provided that the system can be treated as a low-density gas of quasiparticles, or equivalently, that the system is deep within the superfluid, with only weak thermal and quantum fluctuations.

This criterion can be quantified by calculating the condensate depletion, the quasiparticle contribution to the total particle density. The leading contribution is the mean-field term due to the condensate, $\sum_{\ell\gamma} |A_{\ell\gamma}|^2$, while the first-order correction includes both thermally excited quasiparticles and zero-point quantum fluctuations, integrated over $\BZ$. We define the depletion as the ratio of this correction to the mean-field result.

For nonzero temperature $T$, the momentum integral in fact diverges logarithmically, an instance of the Mermin-Wagner-Hohenberg theorem \cite{Mermin,Hohenberg}, which forbids breaking of a continuous symmetry in two dimensions for $T>0$. In the infinite system, there is therefore no true condensate, and so the `depletion' is complete. In a real system, the small-momentum divergence is cut off at a scale $k_0$, given either by the finite size $R\sub{eff}$ of the system in the two-dimensional plane ($k_0 \approx R\sub{eff}^{-1}$) or by a small hopping matrix element $t_{\perp}$ in the transverse direction ($k_0 \approx \sqrt{2m^* t_{\perp}}$, where $m^*$ is the effective mass in the lowest band).

\begin{figure}
\putinscaledfigure{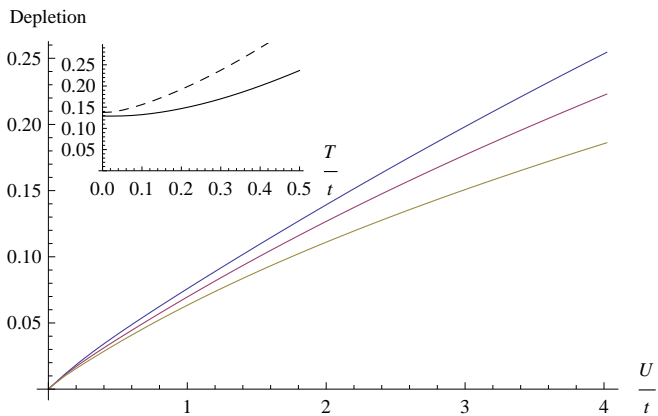}
\caption{\label{FigDepletion}(Color online) Condensate depletion for $\alpha = \frac{1}{3}$ as a function of interaction $U/t$ (main figure) and temperature $T/t$ (inset), where $t$ is the hopping strength. In the main figure, $T = 0$ and the densities are $\rho = 1$ (top curve), $2$ (middle), and $4$ (bottom); in the inset, $\rho = 1$ and $U/t = 2$. The depletion is smallest, and hence the approximation best, deep in the superfluid phase, with weak interactions, high density, and low temperature. For $T>0$, the logarithmic divergence of the depletion integral is removed with small-momentum cutoffs of $k_0 = 0.1$ (solid line) and $0.02$ (dashed line) lattice units, corresponding to an effective system radius of $10$--$50$ sites.}
\end{figure}
Figure~\ref{FigDepletion} shows the depletion for $\alpha = \frac{1}{3}$, as a function of density, interaction strength, and (in the inset) temperature. It is small deep within the superfluid phase and increases to roughly $25\%$ for the largest values of $U$ and $T$ shown. Neglecting cubic and quartic terms within the Bogoliubov theory relies on the assumption of small depletion, and so the conclusions presented here are only qualitatively applicable for larger values of $U$ and $T$.

The depletion calculation also provides a rough estimate for the boundary of the superfluid phase, at the point where the depletion reaches $100\%$, although the approximation of independent quasiparticles is probably not valid at this point. For $\alpha = \frac{1}{3}$, $\rho = 1$ and $T = 0$, this gives an estimate of $(t/U)\sub{c} = 0.08$, in reasonable agreement with the value of $(t/U)\sub{c} = 0.063$ (at the tip of the $\rho = 1$ Mott lobe) found using the Gutzwiller ansatz \cite{Umucalilar}. It should be noted that the latter approach, which neglects fluctuations within the Mott insulator, generally underestimates $(t/U)\sub{c}$ \cite{CapogrossoSansone}.


Techniques to measure the quasiparticle spectrum in ultracold atomic experiments \cite{Stewart,Kollath} include Bragg spectroscopy \cite{Stenger,Rey}, where a two-photon Raman transition is used to measure the dynamic structure factor, a four-point correlation function of the boson operators $b_{\kv}$. In the presence of a condensate, the dominant contribution factorizes into the product of the condensate density and the two-point correlation function of the quasiparticle operators $d_{\kv\zeta}$. The structure factor at frequency $\omega$ is therefore given by a delta function $\delta(\omega \pm \xi_{\kv \zeta})$ at the energy of each quasiparticle mode $\zeta$, allowing the quasiparticle spectrum to be measured directly.


We have presented a theory describing the superfluid phase of bosons in a magnetic field on a square lattice, and showed how many-body effects modify the Hofstadter spectrum of noninteracting particles. Our theory describes the spatial symmetry breaking of the condensate wavefunction and allows the quasiparticle spectrum to be found. We predict clear signatures for time-of-flight images and Bragg spectroscopy, which we expect to be observable in experiments with ultracold atoms in the near future.

We thank Ian Spielman and Trey Porto for helpful discussions. This work is supported by JQI-NSF-PFC and ARO-DARPA-OLE.

\newcommand{\PRA}[3]{Phys.\ Rev.\ A {\bf #1}, #2 (#3).}
\newcommand{\PRB}[3]{Phys.\ Rev.\ B {\bf #1}, #2 (#3).}
\newcommand{\PRL}[3]{Phys.\ Rev.\ Lett.\ {\bf #1}, #2 (#3).}


\begin{thebibliography}{99}

\bibitem{Harper} P. G. Harper, Proc.\ Phys.\ Soc.\ A {\bf 68}, 874 (1955).

\bibitem{Zak} J. Zak, Phys.\ Rev.\ {\bf 134}, A1602; {\bf 134}, A1607 (1964).

\bibitem{Hofstadter} D. Hofstadter, \PRB{14}{2239}{1976}

\bibitem{Jaksch} D. Jaksch and P. Zoller, New J. Phys. {\bf 5}, 56 (2003).

\bibitem{Mueller} E. J. Mueller, Phys.\ Rev.\ A {\bf 70}, 041603(R) (2004).

\bibitem{Sorensen} A. S. S\o rensen et al., Phys.\ Rev.\ Lett.\ {\bf 94}, 086803 (2005).

\bibitem{Gerbier} F. Gerbier and J. Dalibard, New J. Phys.\ {\bf 12}, 033007 (2010).

\bibitem{Schweikhard} V. Schweikhard et al., Phys.\ Rev.\ Lett.\ {\bf 99}, 030401 (2007).

\bibitem{Williams} R. A. Williams et al., Phys.\ Rev.\ Lett.\ {\bf 104}, 050404 (2010).

\bibitem{Cooper} N. R. Cooper, Adv.\ Phys.\ {\bf 57}, 539 (2008).

\bibitem{Spielman} I. B. Spielman, Phys.\ Rev.\ A {\bf 79}, 063613 (2009).

\bibitem{Lin} Y.-J. Lin et al., Phys.\ Rev.\ Lett.\ {\bf 102}, 130401 (2009); Nature {\bf 462}, 628 (2009).

\bibitem{Reijnders} J. W. Reijnders and R. A. Duine, Phys.\ Rev.\ Lett.\ {\bf 93}, 060401 (2004); Phys.\ Rev.\ A {\bf 71}, 063607 (2005).

\bibitem{Goldbaum} D. S. Goldbaum and E. J. Mueller, Phys.\ Rev.\ A {\bf 77}, 033629 (2008); Phys.\ Rev.\ A {\bf 79}, 063625.

\bibitem{Niemeyer} M. Niemeyer et al., \PRB{60}{2357}{1999}

\bibitem{Oktel} M. \"O.\ Oktel et al., Phys.\ Rev.\ B {\bf 75}, 045133 (2007).

\bibitem{Umucalilar} R. O. Umucal\i lar and M. \"O.\ Oktel, \PRA{76}{055601}{2007}

\bibitem{Sinha} S. Sinha and K. Sengupta, arXiv:1003.0258v1 (unpublished).

\bibitem{Bogoliubov} N. N. Bogoliubov, J. Phys.\ (USSR) {\bf 11}, 23 (1947).

\bibitem{AGD} A. A. Abrikosov, L. P. Gorkov, and I. E. Dzyaloshinski, {\it Methods of Quantum Field Theory in Statistical Physics} (Dover, New York, 1963).


\bibitem{Teitel} S. Teitel and C. Jayaprakash, \PRL{51}{1999}{1983}

\bibitem{Polini} M. Polini et al., \PRL{95}{010401}{2005}

\bibitem{Kasamatsu} K. Kasamatsu, Phys.\ Rev.\ A {\bf 79}, 021604(R) (2009).

\bibitem{Lim} L.-K. Lim et al., Phys.\ Rev.\ Lett.\ {\bf 100}, 130402 (2008); Phys.\ Rev.\ A {\bf 81}, 023404 (2010).

\bibitem{Zhai} H. Zhai et al., Phys.\ Rev.\ Lett.\ {\bf 104}, 145301 (2010).

\bibitem{Greiner} M. Greiner et al., Nature {\bf 415}, 39 (2002).

\bibitem{Jaksch1} D. Jaksch et al., \PRL{81}{3108}{1998}

\bibitem{Balents} L. Balents et al., \PRB{71}{144508}{2005}

\bibitem{Blaizot} J.-P. Blaizot and G. Ripka, {\em Quantum theory of finite systems} (MIT Press, Cambridge, Mass., 1986).

\bibitem{Mermin} N. D. Mermin and H. Wagner, Phys.\ Rev.\ Lett.\ {\bf 17}, 1133 (1966).
\bibitem{Hohenberg} P. C. Hohenberg, Phys.\ Rev.\ {\bf 158}, 383 (1967).

\bibitem{CapogrossoSansone} B. Capogrosso-Sansone et al., \PRA{77}{015602}{2008}

\bibitem{Stewart} J. T. Stewart et al., Nature {\bf 454}, 744 (2008).

\bibitem{Kollath} C. Kollath et al., \PRA{76}{063602}{2007}

\bibitem{Stenger} J. Stenger et al., Phys.\ Rev.\ Lett.\ {\bf 82}, 4569 (1999).

\bibitem{Rey} A. M. Rey et al., \PRA{72}{023407}{2005}

\end{thebibliography}
\end{document}